\documentstyle[prb,aps,twocolumn,psfig,floats]{revtex}
\begin{document}
\draft
\twocolumn[\hsize\textwidth\columnwidth\hsize\csname
@twocolumnfalse\endcsname
\title{Superconducting and Magnetic Properties of 
Nb/Pd$_{1-x}$Fe$_x$/Nb Triple Layers}
\author{ Maya Sch\"ock, Christoph S\"urgers, Hilbert v. L\"ohneysen}
\address{Physikalisches Institut, Universit\"at Karlsruhe, D-76128 Karlsruhe,
Germany}
\maketitle
\begin{abstract}
The superconducting and magnetic properties 
of Nb/Pd$_{1-x}$Fe$_x$/Nb triple layers 
with constant Nb layer thickness 
$d_{\rm Nb}$ = 200 \AA\, and different interlayer thicknesses 
3 \AA\,$\leq d_{\rm PdFe} < 80$\,\AA\, are investigated. 
The thickness dependence of the magnetization and of the superconducting 
transition temperature shows that for small iron concentration 
$x$ the Pd$_{1-x}$Fe$_x$\, layer  
is likely to be in the paramagnetic state for very thin films 
whereas ferromagnetic order is established 
for $x \geq 0.13$.  
The parallel critical 
field $B_{c2 \parallel}(T)$ exhibits a crossover 
from two-dimensional (2D) behavior 
where the Nb films are 
coupled across the interlayer, towards a 2D behavior of 
decoupled Nb films with increasing $d_{\rm PdFe}$ and/or $x$.  
This 2D-2D crossover allows a determination of 
the penetration depth $\xi_F$ of Cooper pairs into the 
Pd$_{1-x}$Fe$_x$\, layer 
as a function of $x$. For samples with a ferromagnetic interlayer  
$\xi_F$ is found to be independent of $x$. 
\end{abstract}
]

\section{Introduction}
The proximity effect of a superconductor ($S$) in contact with 
a ferromagnet ($F$) has attracted considerable new interest, since an oscillatory 
behavior of the Cooper pair amplitude in the ferromagnet was 
predicted theoretically in $S/F$\, multilayers \cite{buzd90,rado91}. 
Due to the exchange field in the ferromagnet, the pair-breaking parameter 
is complex and causes a spacial modulation 
of the superconducting order parameter in the ferromagnetic interlayer. 
For certain thicknesses $d_F$ 
of the ferromagnetic layer 
the phase of the order parameter changes by 
$\Delta \phi = \pi$
across the barrier (so-called $\pi$-junction \cite{bula77})
which gives rise to an enhanced transition temperature $T_c$. 
This, for instance, 
should show up in a nonmonotonic dependence of $T_c(d_F)$.  
Further theoretical work has shown 
that these anomalies should also 
occur in $S/F$ bilayers \cite{deml97,khus97}.
Several experimental studies have been performed to search for the 
appearance of $\pi$-coupling in $S/F$ multilayers and triple layers
\cite{wong86,stru94,koor94,koor95,jian95,jian96,merc96}. 
However, up to now 
none of the investigations have revealed unambiguously a nonmonotonic 
behavior due to a $\pi$-coupling mechanism. 
Although the measurements on sputtered Nb/Gd multilayers 
and triple layers have been 
interpreted in terms of this mechanism \cite{jian95,jian96}, 
the loss of ferromagnetic order at thin interlayer 
thicknesses \cite{stru94} or a magnetically "dead" interface 
region can also result in 
a nonmonotonic behavior of $T_c(d_F)$ \cite{mueh96}. 
Furthermore, the magnitude of the 
electron mean-free path $l$ in $S$ and $F$, the interface transparency, and 
spin-orbit scattering must be taken into account 
\cite{deml97,khus97,gari98}.
A prominent parameter which enters all of the present theories 
is the characteristic complex decay constant $k_F$ describing the 
decay of the pair amplitude $F_F$ in the $F$\, layer along the 
surface normal $x$, i.e. $F_F \propto 
exp(-k_F x)$. The real part of $k_F$ defines the exponential 
decay of the envelope of $F_F$, i.e. the penetration depth of Cooper 
pairs in the $F$ layer, whereas the imaginary part defines oscillations 
of $F_F$. In the theory of Radovi\'{c} et al. these two length scales turn 
out to be identical, $({\rm Im} k_F)^{-1} = ({\rm Re} k_F)^{-1} = \xi_F/2$, 
with the characteristic length $\xi_F$ defined as 
$\xi_F = \sqrt{4 \hbar D_F /I}$.
 $D_F$ is the electronic diffusion constant in $F$ and $2I$ is the splitting 
of the spin-up and spin-down conduction bands by the exchange interaction 
\cite{rado91}.  
However, the penetration depth and the oscillation period of $F_F$ 
can be different for small electron mean free paths and strong 
spin-orbit scattering \cite{deml97,khus97}. 
In recent experiments which have been mainly discussed in frame 
of the theory by Radovi\'{c} et al. the length $\xi_F$ often
serves as an adjustable parameter. 
In this paper, we will focus on the upper critical field of 
triple layers and show that $\xi_F$\, can be determined 
from a crossover in the parallel upper critical field 
of Nb/Pd$_{1-x}$Fe$_x$/Nb 
triple layers rather than use it as a free parameter
whose value depends on the theoretical model employed. 
This crossover 
from two-dimensional (2D) behavior 
of the whole triple layer to 2D behavior
of each Nb film individually, 
can only be observed for a thickness $d_F$\,
smaller than a critical thickness $d_c$, 
which will be identified as $\xi_F$, see below. 
The influence of different ferromagnetic 
materials on the superconducting properties of $S/F$\, multilayers 
has been studied previously by Koorevvar 
et al. \cite{koor95} where the emphasis was put on the 
critical thickness of the superconducting layers which are decoupled by
thick ferromagnetic interlayers. In contrast, the present work focuses on 
the influence of the ferromagnetic layer thickness in $S/F/S$\, triple layers 
with superconducting layers of constant thickness. We will furthermore show that 
the analysis of the superconducting properties allows  
access to the magnetic properties in thin ferromagnetic films.
The ferromagnetic behavior of Pd$_{1-x}$Fe$_x$\, alloys has been studied 
in great detail, in particular the Pd-rich alloys \cite{nieu75}. 
In Pd$_{1-x}$Fe$_x$, the Curie temperature $T_{\rm C}$\, can be changed over a wide 
range of concentration $x$. For small $x$, the polarization 
of the Pd conduction band around each localized 
Fe moment gives rise to a "giant moment" 
of 13 - 16 $\mu_B$\, per Fe impurity \cite{herrm96}. 
In addition, the Fe-Pd 
exchange interaction leads to an indirect ferromagnetic Fe-Fe interaction.   
In bulk alloys, ferromagnetism persists down to 
a concentration of about $x \approx 10^{-4}$ (Ref. \onlinecite{bues92}) 
and $T_{\rm C}$\, increases monotonically with increasing $x$.  
Furthermore, with increasing $x$ 
the electronic structure gradually changes from a 
localized to a more itinerant behavior \cite{mohn93} and direct Fe-Fe 
interactions become important. In the present work we use the large variability 
in $T_{\rm C}$\, and interaction strength with $x$ to tune the 
$S/F$\, coupling.
 
\section{Experimental}
Nb single layers and Nb/Pd$_{1-x}$Fe$_x$/Nb triple layers 
were grown in an ultra-high vacuum system 
by e-beam evaporation onto 
in-situ cleaned $\rm{Al_{2}O_{3}(11\bar20)}$\ substrates 
at room temperature.
A set of eight samples with different $d_{\rm PdFe}$ was prepared 
during a single evaporation process. 
For the triple layers, a 200-\AA\ Nb film was 
deposited first onto all substrates of one set. 
The thicknesses were measured 
with a quartz-crystal monitor. 
During the subsequent simultaneous evaporation of 
Pd and Fe from two different
crucibles the computer-controlled sample shutter 
was opened stepwise to expose the samples 
one after another to the Pd and Fe beams. 
Finally, a second 200-\AA\ Nb layer was deposited on all samples.
The relative error of the Fe concentration estimated from the variation of 
the evaporation rates during the process was less than 10 \%.
Symmetrical $\theta/2 \theta$ scans were taken on some samples 
with a thick Pd$_{1-x}$Fe$_x$ interlayer ($d_{\rm PdFe} >$\,30 \AA)
using a standard X-ray powder diffractometer with Cu-$K_{\alpha}$ radiation. 
The scans indicate oriented growth of bcc-Nb (110) and fcc-PdFe (111) 
for $x \leq 0.4$\, or bcc-Fe for $x = 1$\, along the surface normal. 
The measured lattice parameters agree within 10 \%  
with data of 1000-\AA\, rf-sputtered 
Pd$_{1-x}$Fe$_x$ films \cite{zhan88}. 
   
The electrical resistivity was measured in a 
$\rm{He}^4$ cryostat with a 
conventional four-point probe using 
spring-loaded needles in magnetic 
fields up to 5 T. The superconducting critical temperature 
$T_c$ was determined from the midpoint of the resistive transition. 
The transition width $\Delta T_c$, estimated from 
the difference in temperature 
at 10 \% and 90 \% of the transition, was typically $\Delta T_c \leq$\,40 mK 
for all samples.
The upper critical magnetic field $B_{c2}(T)$ 
was determined from the midpoint
of the transition measured by ramping up the magnetic field until
90\% of the transition was accomplished and ramping 
down again. The temperature during this sweep was held 
constant within typically 
2 mK. The critical-field transition width is defined as the field difference 
at 10 \% and 90 \% of the transition. 
The magnetic properties were investigated by
SQUID magnetometry for temperatures 2 - 300 K and in magnetic 
fields up to 5 Tesla with the field oriented parallel to the sample surface.

\section{Results}
\subsection{Magnetic Properties}
\begin{figure}
\centerline{\psfig{file=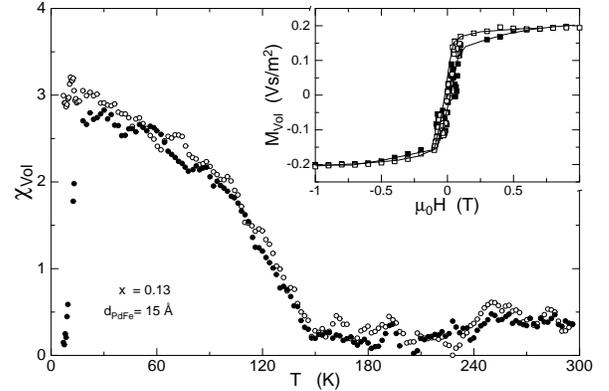,angle=-90,width=8cm}}
\caption[]{Magnetic susceptibility $\chi$ 
vs. temperature $T$ for $x$ = 0.13 and $d_{\rm PdFe}$ = 15 \AA\, 
measured in zero-field cooled (open circles) 
and field-cooled mode (closed circles). 
The applied magnetic field $B = \mu_0H$ = 10 mT was oriented parallel to the 
film plane.
The inset shows a hysteresis loop $M(H)$ 
taken at $T$ = 8 K (up sweep: solid symbols, down sweep: 
open symbols).}
\label{fig1}
\end{figure}

Fig. \ref{fig1} shows the dc-susceptibility $\chi (T)$ for one triple layer 
measured in the zero-field-cooled and field-cooled modes.
 The sharp decrease of the signal at low $T$ is due to the superconducting 
transition. The Curie temperature $T_{\rm C}$\, was estimated from 
the extrapolation to $M(T \rightarrow T_{\rm C}) = 0$. $T_{\rm C}$\, 
is listed in Table I for some samples together with the
values for bulk Pd$_{1-x}$Fe$_x$ alloys ($d_{\rm PdFe} = \infty$).
 The ferromagnetic order was further checked on several samples of 
different concentration and Pd$_{1-x}$Fe$_x$ thickness by 
performing hysteresis 
loops $M(H)$ at $T = 8$\,K (Fig. \ref{fig1}, inset).

The magnetic properties of the films change with  
thickness {\it and} concentration.
Table I clearly shows that $T_{\rm C}$\, 
decreases with decreasing $x$ when samples 
of almost equal $d_{\rm PdFe}$ 
are compared, as expected from the $T_{\rm C}(x)$-dependence 
in the respective bulk alloys \cite{nieu75}. 
Furthermore, $T_{\rm C}$\, decreases 
with decreasing $d_{\rm PdFe}$\, for fixed concentration, e.g. $x = 0.20$. 
This is possibly due to finite-size effects where the thickness 
dependence of the Curie temperature is 
described by $T_{\rm C} \propto d_F^{-\lambda}$ and 
the exponent $\lambda$ depends on the dimensionality and universality 
class of the system. 
For $x = 0.05$ the magnetic signal was very weak (not shown). 
From the susceptibility data a $T_{\rm C} \approx 60$\,K was estimated. 
For $x = 0.01$ a magnetic signal could not be 
detected. However, the magnetic behavior of these samples can be inferred 
from the investigation of the superconducting 
properties as will be discussed below.
The magnetic moment per atom $\mu_{\rm exp}$\, 
was determined from the saturation
magnetization (Table I), albeit with a large error for
small $d_{\rm PdFe}$. The average moment increases with 
concentration $x$ for samples with roughly 
equal thickness, in accordance with the concentration dependence 
$T_{\rm C}(x)$ in bulk alloys. 
However, the measured values 
are smaller than those of
bulk samples ($\mu_{\rm exp} / \mu_{\rm bulk} < 1$, Table I). 
This could be due to the existence of a magnetically "dead" layer 
at the Pd$_{1-x}$Fe$_x$/Nb interface as reported previously for 
Fe/Nb/Fe triple layers \cite{mueh96}. 
In this case, the measured values can be described by 
$\mu_{\rm exp} / \mu_{\rm bulk} = 1-d_0/d$, where $d_0$ is the thickness 
of the nonmagnetic layer. From the data of Table I a thickness 
$d_0 \approx$\,8 \AA\, is estimated, i.e. 4 \AA\, on either 
side of the Pd$_{1-x}$Fe$_x$\, layer. 
The latter is smaller than the interface thickness of $\approx 7$\,
\AA\, between Nb and pure Fe \cite{mueh96}.

\begin{table}
\caption {Magnetic properties of Nb/Pd$_{1-x}$Fe$_x$/Nb triple layers}
\begin{tabular}{ccccccc}
$ x $\ & $d_{\rm PdFe}$ (\AA)\ & $T_{\rm C}$ (K)\ & $\mu_{\rm exp} (\mu_{\rm B})$\ 
& $\mu_{\rm exp} / \mu_{\rm bulk}\ $ & Ref. \\
\hline
0.05 & 18 &  60 & $0.17 \pm 0.1$ & $0.57 \pm 0.3$ & this work  \\
   & $\infty$ & 162 & 0.3 & 1 & \onlinecite{cran60} \\
0.13 & 15 & 150 & $0.2 \pm 0.1$ & $0.3 \pm 0.2$ & this work \\
   & $\infty$ & 320 & 0.66 & 1 & \onlinecite{cran60} \\
0.20 & 9 & $ 70$ & $0.13 \pm 0.1$ & $0.14 \pm 0.1$ & this work \\
   & 12 & 175 & $ 0.1 \pm 0.1$ & $0.1 \pm 0.1$ & this work \\
   & 17 & $> 300$ & $0.96 \pm 0.24$ & $1.05 \pm 0.4$ & this work \\
   & $\infty$ & 440 & 0.91 & 1 & \onlinecite{cran60} \end{tabular}
\end{table}

\subsection{Upper Critical Magnetic Field of Nb Single Films}

Single Nb films with various thicknesses were investigated 
in order to check that the parallel critical field of 
Nb/Pd$_{1-x}$Fe$_x$/Nb triple layers is not determined by the 
occurrence of surface superconductivity. 
Before presenting the results, 
we briefly summarize the usual behavior of $B_{c2} (T)$ in dependence 
of the orientation of the magnetic field and of the dimensionality of the 
sample. 
In general, the perpendicular critical 
magnetic field of superconducting films of thickness $d$ 
obeys a linear temperature dependence, 
i.e. three-dimensional (3D) behavior, below $T_c$, 
\begin{equation}
B_{c2 \perp}(T) = \frac{\phi_0}{2 \pi \xi_{0\parallel}^2} (1-T/T_c)
\label{b3perp}
\end{equation}
because the sample dimensions are much larger than 
the temperature dependent Ginzburg-Landau coherence 
length parallel to the film plane,
$\xi_{\parallel}(T) = \xi_{0\parallel}/
\sqrt{1-T/T_c}$, with $\xi_{0\parallel} = \xi_{\parallel}(T=0)$. 

In the parallel orientation, $B_{c2 \parallel} (T)$ can be described 
by a similar expression where $\xi_{0\parallel}^2$ is 
replaced by $\xi_{0\parallel} \xi_{0\perp}$.   
Even in isotropic superconductors such as Nb, a difference in 
$\xi_{0\parallel}$ and $\xi_{0\perp}$ may occur because of an anisotropic 
microstructure. If just below $T_c$ 
the perpendicular coherence 
length $\xi_{\perp}$ is larger than the thickness for very thin films,
$ \xi_{\perp}(T) \ >> d$, the temperature dependence of 
$B_{c2 \parallel}$ is described by the Tinkham expression for 
two-dimensional (2D) superconductors \cite{tink75},
\begin{equation}
B_{c2 \parallel}(T) = \frac{\sqrt{12} \phi_0} 
{2 \pi \xi_{0\parallel} d} \sqrt{1-T/T_c}.
\label{b2para}
\end{equation}
Apart from a factor $\sqrt{12} /\pi$ the same result is obtained 
by using a Ginzburg-Landau 
approach for dirty and anisotropic superconductors, as has been done by 
Schneider and Locquet \cite{schn91} 
to describe the overall temperature dependence 
of $B_{c2 \parallel}(T)$. 
At lower temperatures $\xi_{\perp} (T)$ 
can become smaller than the film thickness  
and the 3D behavior is recovered.
Hence, because of the temperature dependence of $\xi(T)$ 
a dimensional crossover from 2D to 3D behavior 
should occur in $B_{c2 \parallel}(T)$ 
with decreasing $T$.
 However, for sufficiently thick films 
this regime will not be entered 
due to the onset of surface superconductivity which occurs 
for $1.84 \xi_{\perp}(T) < d$, i.e. when the 
diameter of a vortex $\approx 2\, \xi (T)$ is smaller than the film thickness 
\cite{fink69}. 
In a field decreasing from well above $B_{c2 \parallel}$, 
nucleation of superconducting regions will start near the surface 
leading to a superconducting sheath for fields $B_{c2 \parallel} 
< B < B_{c3 \parallel} = 1.69\, B_{c2 \parallel}$ \cite{sain63}. 
Generalizing this to anisotropic superconductors one obtains a 
linear behavior     
\begin{equation}
B_{c3 \parallel}(T) = \frac{1.69 \phi_0}{2 \pi 
\xi_{0\parallel} \xi_{\perp}} (1-T/T_c).
\label{b3surf}
\end{equation}

Therefore, in superconducting 
films a dimensional crossover 
from 2D (Eq. \ref{b2para}) to surface superconductivity 
(Eq. \ref{b3surf}) can occur as a function 
of temperature and thickness in the parallel critical field.
 
\begin{figure}[t!]
\centerline{\psfig{file=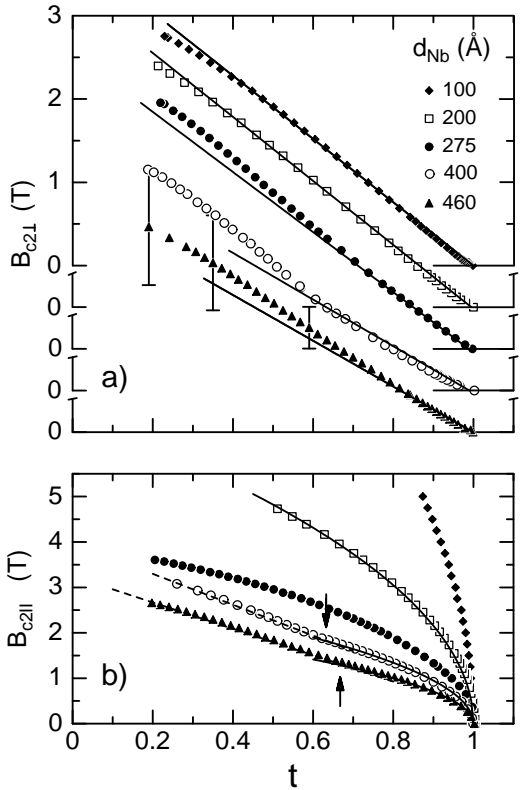,angle=-90,width=8cm}}
\caption[]{(a) Perpendicular critical magnetic field $B_{c2 \perp}$ vs. 
reduced temperature $t = T/T_c$ for Nb single films of different thickness. 
Solid lines indicate the linear temperature dependence near $t = 1$. 
The error bars do not exceed the symbol size for $t \geq 0.7$. 
(b) Parallel critical field $B_{c2 \parallel} (t)$. Arrows indicate the 
temperature $t^{\ast}$ below which surface superconductivity comes into play. 
Solid and dashed lines show the square-root and linear behavior above and 
below $t^{\ast}$, respectively.}
\label{fig2}
\end{figure}

Fig. \ref{fig2}a shows the temperature dependence of $B_{c2 \perp}$ 
for single films with different thickness 
$d_{\rm Nb}$ vs. the reduced transition temperature $t = T/T_c$. 
All films show a 
linear dependence for $t > 0.75$\, characteristic for 
3D behavior (Eq. \ref{b3perp}).
For lower temperatures, the data points of films with 
$d_{\rm Nb} \geq$\, 275 \AA \, deviate from the linear dependence with 
a concomitant increase in the transition width. $\Delta T_c$ 
is indicated by vertical bars. This behavior is presumably 
caused by thermally activated flux creep, 
which is more likely to occur in samples 
with a lower concentration of pinning centers. 
This is the case for thicker films 
which have a lower resistivity (Table II), 
i.e. a lower concentration of defects.

\begin{figure}
\centerline{\psfig{file=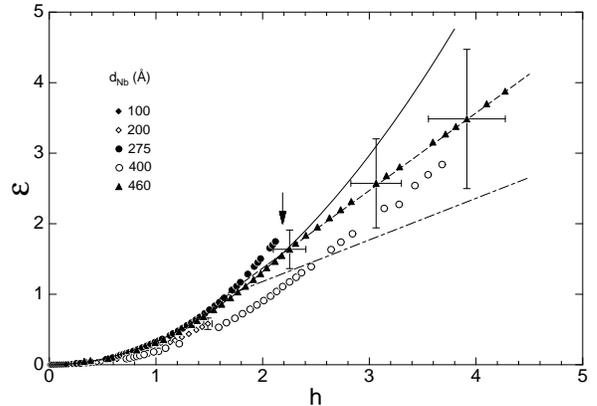,angle=-90,width=8cm}}
\caption[]{Plot of the scaled perpendicular critical field 
$\epsilon = B_{c2 \perp} (\pi d^2)/(2 \phi_0)$ 
vs. the scaled parallel critical field $h = B_{c2 \parallel} 
(\pi d^2)/(2 \phi_0)$. 
Solid and dashed lines indicate the experimental behavior 
for the 460-\AA\, film
$\epsilon \propto 0.33 h^2$ and $\epsilon \propto 1.1 h$, respectively. 
The transition takes place around $h \approx 2.2$ marked by arrow. 
Dashed-dotted line 
indicates $\epsilon = h/1.695$ without anisotropy $(\xi_{0 \parallel} = 
\xi_{0 \perp})$. Horizontal and vertical bars indicate the transition widths.}
\label{fig3}
\end{figure}

The parallel critical field $B_{c2 \parallel} (t)$ is 
shown in Fig. \ref{fig2}b. 
For $t > 0.7$ all samples show a square-root dependence of $B_{c2 \parallel}$, 
i.e. 2D behavior (Eq. \ref{b2para}), 
indicating that $\xi_{\perp}(t)$ is larger than the 
thickness $d_{\rm Nb}$. 
For films with $d_{\rm Nb} \leq 275$\,\AA\, 
the 2D behavior survives down to the lowest temperature. 
In contrast, in thicker films with $d_{\rm Nb} \geq 400$\,\AA,  
where below a certain temperature $t^{\ast}$\,
the nucleation center moves from the 
center of the film to the surface, the temperature dependence 
of $B_{c2}(t)$ changes from square-root to linear behavior due to the onset 
of surface superconductivity.  
The reduced perpendicular and parallel 
critical fields, i.e. $\epsilon =  B_{c2 \perp}(T) d^2 \pi/2 \phi_0$  and 
$h = B_{c2 \parallel}(T) d^2 \pi / 2 \phi_0$, are plotted 
in Fig. \ref{fig3} with $T$ as an implicit parameter. 
For small $h$ the data follow a single line $\epsilon = 0.33 h^2$ 
in agreement with the theoretical prediction 
for the 2D regime according to Saint-James and de Gennes \cite{sain63}.
In films with $d \geq$\, 400 \AA\, a change to a linear 
behavior due to the superconducting surface sheath is observed for 
$h > 2.2$, e.g. $\epsilon \approx 1.1 h - 0.82$ for the 460-\AA\, film. 
A behavior $\epsilon = (\xi_{0\perp} / 1.695\,\xi_{0\parallel}) h$ 
is expected when the anisotropy of the 
coherence length (Eqs. \ref{b3perp} and \ref{b3surf}) is taken into account.
The observed offset from a strict proportionality $\epsilon \propto 
h$\, reflects the slightly superlinear $t$\, dependence of $B_{c2 \perp}$. 
We neglect this fact for the following qualitative discussion. 
The crossover between the two regimes at $h \approx 2.2$ 
corresponds to the reduced temperature $t^{\ast} = T^{\ast}/T_c$ 
marked by arrows in Fig. \ref{fig2}b. 
We obtain an anisotropy ratio $\xi_{0\perp}/
\xi_{0\parallel}$ about 2 from the linear regime. 
With $\xi_{0\parallel}$ determined from $B_{c2 \perp}$, the coherence 
length $\xi_{0\perp}$ can be derived (see Table II). 

The fact that $\xi_{0\perp}$ 
is larger than $\xi_{0\parallel}$ 
can be explained by the anisotropic microstructure of the film, 
with columns perpendicular to the surface in addition to fine 
equiaxed grains, giving rise to an anisotropic electron diffusivity
 \cite{suer94}. Alternatively, in the dirty limit an average coherence length 
$\xi_0^{\rho}$ can be determined from electronic mean free path $l$ \cite{stru94},
which was calculated from the residual resistivity $\rho$ using 
$\rho l = 3.75 \times 10^{-16} \Omega {\rm m}^2$\, \cite{webe91}. 
 It is reassuring that these values lie between the 
values of $\xi_{0\parallel}$ and $\xi_{0\perp}$ (see Table II). 
We conclude that for temperatures $T \geq$\,1.5 K investigated here, 
Nb films with $d_{\rm Nb} \leq 275$ \AA\, show 
only 2D behavior in the parallel critical field 
without the occurrence of surface superconductivity, whereas in thicker films
$B_{c2 \parallel} (T)$ changes due to the onset of surface superconductivity. 
In this case, $\xi_{0\perp}$ can be measured. This is important 
for the following discussion of Nb/Pd$_{1-x}$Fe$_x$/Nb triple layers with 
$d_{\rm Nb}$ = 200 \AA. 
Moreover, we note that 
in Nb/Pd$_{1-x}$Fe$_x$/Nb triple layers surface 
superconductivity is likely to occur only 
at even larger thicknesses due to 
the pairbreaking at the $S/F$ interface.      

\begin{table}[h!]
\caption {Superconducting parameters of Nb films}
\begin{tabular}{ccccccc}
$d_{\rm Nb}$\ & $\rho (\mu \Omega {\rm cm})$\ & $l$\,(\AA) \ & $\xi_0^{\rho}$ (\AA) & 
$\xi_{0\parallel}$ (\AA) & $\xi_{0\perp}$ (\AA) & $\xi_{0\perp} / \xi_{0\parallel}$ \\ 
\hline
100 & 15.35 & 24 & 79 & 90 & &\\
200 & 11.86 & 32 & 90 & 91 & &\\
275 & 8.0  & 47 & 107 & 93 & &\\
400 & 5.47 & 69 & 126 & 104 & 182 & 1.75 \\
460 & 5.50 & 68 & 126 & 103 & 210 & 2.04 
\end{tabular}
\end {table}

\subsection{Upper Critical Magnetic Field of 
Nb/Pd$_{1-x}$Fe$_x$/Nb Triple Layers}

\begin{figure}
\centerline{\psfig{file=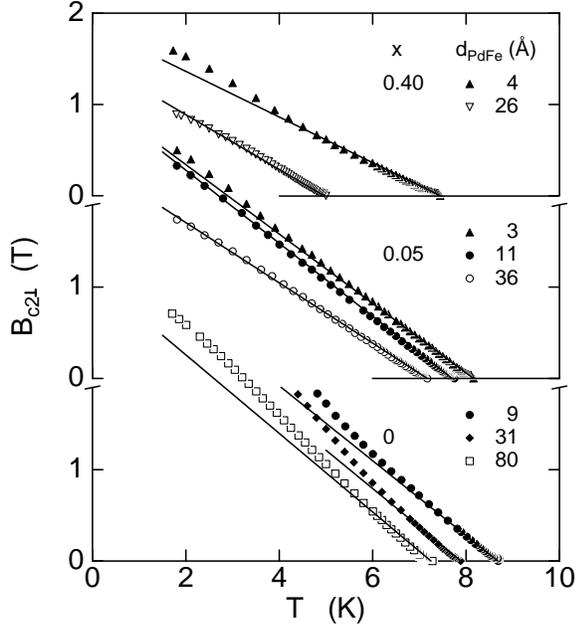,angle=-90,width=8cm}}
\caption[]{Perpendicular critical magnetic field $B_{c2 \perp}$ vs. 
temperature $T$ for Nb/
Pd$_{1-x}$Fe$_x$/Nb triple layers 
of different thickness $d_{\rm PdFe}$ and concentration $x$. 
Solid lines indicate the linear temperature dependence near $T_c$.}
\label{fig4}
\end{figure}

The perpendicular critical magnetic field 
for some selected samples of different 
$d_{\rm PdFe}$ and $x$ is shown in Fig. \ref{fig4}. 
In all cases, an almost linear $T$ dependence suggesting 3D behavior 
is observed. Deviations at 
lower temperatures can be attributed to an increasing transition 
width in magnetic field which aggravates the precise determination 
of $B_{c2 \perp}$. 
\begin{figure}
\centerline{\psfig{file=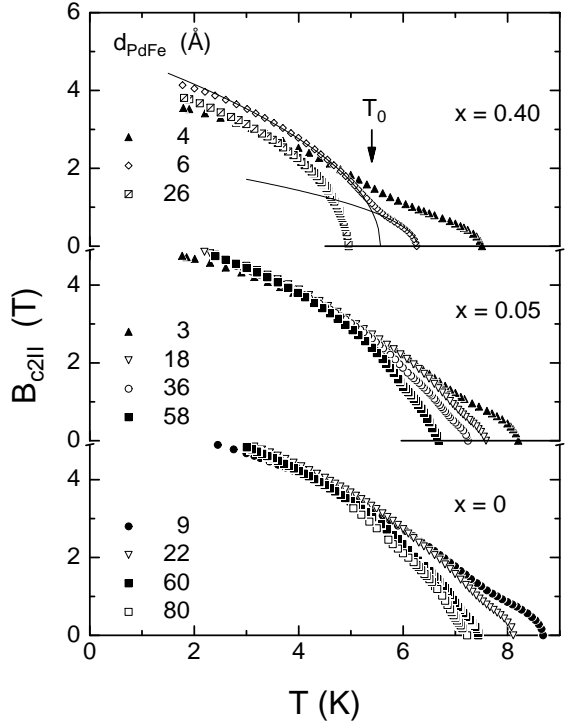,angle=-90,width=8cm}}
\caption[]{Parallel critical magnetic field $B_{c2 \parallel}$ vs. 
temperature $T$ for Pd$_{1-x}$Fe$_x$ triple layers 
of different thickness $d_{\rm PdFe}$ and concentration $x$. 
For $x$ = 0.4 and $d_{\rm PdFe}$ = 6 \AA\, the solid lines 
indicate the different square-root behavior at temperatures below and 
above the 2D-2D crossover temperature $T_0$ marked by arrow.}
\label{fig5}
\end{figure}

\begin{figure}
\centerline{\psfig{file=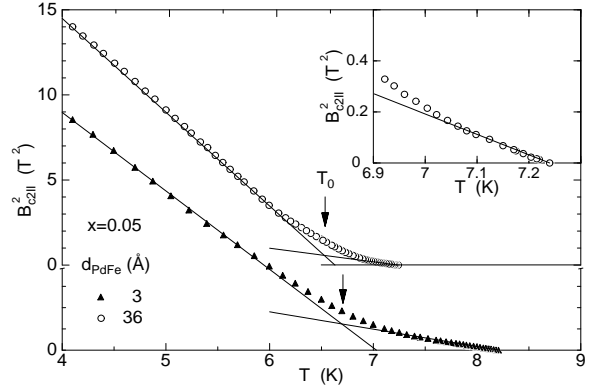,angle=-90,width=8cm}}
\caption[]{$B_{c2 \parallel}^2$ vs. 
$T$\, plot for two triple layers 
of different thickness $d_{\rm PdFe}$ with $x = 0.05$. Solid lines 
indicate the linear behavior in the two regimes. Arrows indicate 
the 2D-2D crossover temperature $T_{0}$. Inset shows the linear behavior 
near $T_c$ for $d_{\rm PdFe}$ = 36 \AA.}
\label{fig6}
\end{figure}

The temperature dependence of 
the parallel critical field $B_{c2 \parallel}$ is shown in Fig. \ref{fig5}. 
Two kinds of behavior are observed:
 
(1) For thin interlayer thickness and/or low iron concentration 
(including $x$ = 0) $B_{c2 \parallel} (T)$ exhibits a square-root like behavior close
to $T_c$ and a second square-root like behavior of 
different slope at lower temperatures. This is illustrated by way of 
example in Fig. \ref{fig6}, 
where $B^2_{c2 \parallel}$ plotted vs. $T$ exhibits two 
linear regimes with a gradual transition around a temperature 
$T_{0}$ indicated by arrows. 
The change in the $T$ dependence at $T_{0}$ and magnetic field $B_{0}$ 
is attributed to a 2D-2D crossover, similar to the case of Pb/Ge multilayers 
with a limited number of bilayers \cite{neer91}.
 Roughly speaking, just below $T_c$ both Nb layers are 
coupled through the interlayer, the order parameter extends over the 
total sample thickness and 
for $\xi_{\perp}(T) > d_{tot}$ a 2D behavior is observed (Eq. \ref{b2para}).
 At lower temperatures the coupling is suppressed by the 
pairbreaking of the interlayer and 
the individual Nb layers give rise to a second 
2D behavior in $B_{c2 \parallel}$. 
The effect of an external magnetic field on the proximity effect 
was studied earlier in $S/N$ junctions \cite{orsa67}. 
In the dirty limit the temperature dependence of the coherence 
length in the nonmagnetic 
metal $N$,  
$\xi_N = \sqrt{D/2 \pi k_{\rm B}}T \sim 1/\sqrt{T}$, 
would suggest that the coupling of the 
$S$ layers increases with decreasing 
temperature. An increasing applied magnetic field
has no effect on the overlap of both pair-amplitudes from $S$\, leaking 
into $N$\, with an exponential decay until 
superconductivity in $N$\, suddenly breaks down at the so-called 
break-down field $B_0$\, \cite{orsa67}. 
At $B_0$\, only a small superconducting 
sheath at the $S/N$\, interface survives, 
which further decreases in thickness with increasing $B$. 
The detailed behavior depends on the thickness $d_N$ and on the 
transition temperature $T_{cN}$ of the $N$ layer.
We simply adopt these results to the investigated 
$S/F$ system,  
assuming that the exchange interaction in $F$ increases the
pairbreaking but does not change the qualitative behavior.
For a strong ferromagnetic interlayer $T_{cN}$ would be zero.
Therefore, the data of samples with a thin and/or weakly 
ferromagnetic Pd$_{1-x}$Fe$_x$-layer can be explained in the following way: 
In low fields the Nb films are coupled and give rise to 
single-film behavior with $d_{tot} \approx 2 d_{\rm Nb} + d_{\rm PdFe}$. 
When the break-down field $B_0$\, is reached, 
superconductivity in the Pd$_{1-x}$Fe$_x$ layer
is almost completely suppressed and the Nb layers decouple, showing a 
2D behavior in $B_{c2 \parallel}$.
(More precisely, the transition in the $(T,B)$ phase diagram 
(Fig. \ref{fig6})
occurs at a point where the $B_0(T)$ line crosses the 
$B_{c2 \parallel}(T)$ curve.) 
This is further confirmed by the calculated coherence lengths 
$\xi_{0 \parallel}$
(Eq. \ref{b2para}) which agree well with those
obtained from the $B_{c2 \perp}(T)$ data if we assume 
$d=d_{tot}$\, for coupled films ($T > T_{0}$) 
and $d = d_{\rm Nb} = 200$\,\AA\, for
decoupled films ($T < T_{0}$). 
The effective thickness of $S$ might be somewhat smaller due to 
the proximity of the ferromagnetic 
material. This would lead to higher 
values of $\xi_{0 \parallel}$\, 
and better agreement with the values determined from the  
$B_{c2 \perp}(T)$ behavior.
We emphasize that the 2D-2D crossover appears at higher 
temperatures 
than $T^{\ast}/T_c \approx 0.6$\,($d$ = 400 \AA) below which the onset of
 surface superconductivity would change the $B_{c2 \parallel}$ behavior 
 as shown above.

\begin{figure}
\centerline{\psfig{file=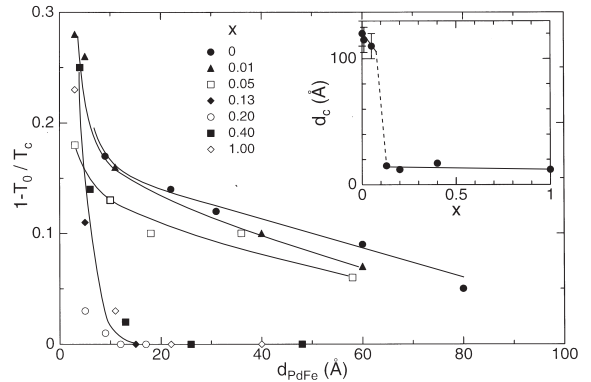,angle=-90,width=8cm}}
\caption[]{Reduced crossover temperature $(1-T_0/T_c)$ vs. $d_{\rm PdFe}$
for different $x$. Solid lines serve as guide to the eye. 
Dashed lines show the extrapolation to $(1-T_0/T_c) = 0$. 
Inset shows the critical thickness $d_c$ derived from the intersection 
of $(1 - T_0 / T_c)$ with the abscissa for different $x$.} 
\label{fig7}
\end{figure}

(2) In samples with a large $d_{\rm PdFe}$ and/or  
large $x$ the individual Nb layers are decoupled for all $T$ 
by the strong pairbreaking of the magnetic interlayer. 
In this case 
a single square-root behavior
of $B_{c2 \parallel}(T)$ is observed 
down to the lowest temperature (cf. Fig. \ref{fig5}, 
$x = 0.40$, $d_{\rm PdFe} = 26$ \AA), 
similar to the case of a single 200-\AA\, Nb film 
(Fig. \ref{fig2}b). 
Hence, the 2D-2D crossover 
depends on the interlayer thickness {\it and} the iron concentration. 
Fig. \ref{fig5} immediately shows that for constant $x$
and increasing $d_{\rm PdFe}$ 
the characteristic temperature $T_0$ shifts towards $T_c$ which in turn
decreases, until the individual Nb films are decoupled for all temperatures 
(cf. Fig. \ref{fig5}, $x = 0.40$). The $T_0(x, d_{\rm PdFe})$ dependence 
is seen more clearly in  
Fig. \ref{fig7} where the reduced values $(1-T_0/T_c)$ 
are plotted vs. $d_{\rm PdFe}$
for different $x$.

The data can be separated into two regimes. 
For small $x \leq 0.05$ a gradual decrease to large $d_{\rm PdFe}$ is 
seen. 
Moreover, for constant $d_{\rm PdFe}$ 
these samples show a systematic decrease of $(1-T_0/T_c)$ 
with increasing $x$. In contrast, for $x \geq 0.13$ 
the data seem to follow a single line with a 
steep decrease to $(1-T_0/T_c) = 0$\, 
at small $d_{\rm PdFe}$. (Note that for $d_{\rm PdFe} \rightarrow 0$, 
$T_0$ should be zero for all $x$.) 
From the intersection of the data with the abscissa 
a critical thickness $d_c$ 
can be determined in dependence of $x$. 
In samples with $x \leq 0.05$ this thickness 
was estimated by a linear extrapolation of the data. 
For each concentration, $d_c(x)$ 
is the smallest $d_{\rm PdFe}$ for which a crossover is still observed 
whereas samples with $d_{\rm PdFe} > d_c$ 
show a $B_{c2 \parallel}(T)$ behavior of decoupled Nb films 
for all temperatures $T < T_c$. 
Thus, in samples with $d_{\rm PdFe} = d_c$ the two pair amplitudes 
decaying from both $S$ layers into the $F$ layer do not overlap, 
i.e. $d_c/2 \approx ({\rm Re} k_F)^{-1} = \xi_F/2$, Ref. \onlinecite{rado91}.
Therefore, in the following $d_c$ will be identified as $\xi_F$ 
as previously done by Koorevar et al. for V/Fe multilayers \cite{koor94}. 
The concentration dependence $d_c(x)$ is plotted in the
inset. Again, the two different regimes can be identified. 
First, $d_c$ decreases 
slightly with increasing $x$ but becomes almost independent of $x$ 
after a precipitous decrease between $x = 0.05$\, and 0.13. 
Obviously, 
these two regimes are related to different 
pairbreaking mechanisms at the $S/F$ interface, which will be discussed 
later. In the following, $d_c$ will be identified as $\xi_F$.

\subsection{Superconducting Transition 
Temperature of Nb/Pd$_{1-x}$Fe$_x$/Nb Triple Layers}

\begin{figure}
\centerline{\psfig{file=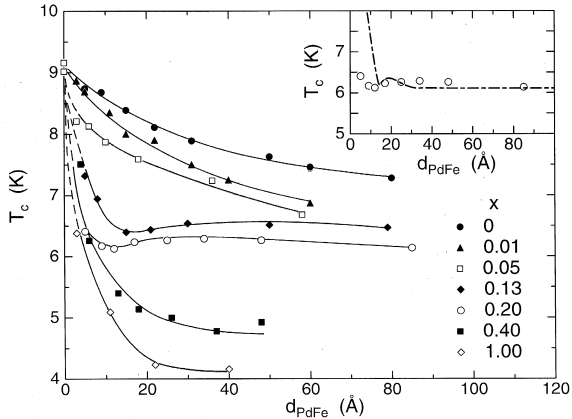,angle=90,width=8cm}}
\caption[]{Superconducting transition temperature $T_c$ vs. 
interlayer thickness 
$d_{\rm PdFe}$ for different $x$. Solid lines serve as guide to the eye. 
Dashed lines show extrapolations to $T_c(d_{\rm PdFe} = 0$). The inset 
shows a fit by the theory of Radovi\'{c} et al. 
\cite{rado91} (dashed-dotted line) 
to the data for $x$ = 0.20, see text for further details.}
\label{fig8}
\end{figure}

The superconducting transition temperatures $T_c$ 
vs. interlayer thickness $d_{\rm PdFe}$ for sample sets of 
different $x$ with constant $d_{\rm Nb}$\, = 200 \AA\, 
are shown in Fig. \ref{fig8}. The transition width 
corresponds to the symbol size.
For pure Pd ($x = 0$) a monotonic decrease of $T_c(d_{\rm PdFe})$ 
from $T_c(0) = 9$\,K is observed. 
Similar behavior has been reported recently for Nb/Pd multilayers 
with $d_{\rm Nb}$ which has been discussed in frame of the 
de Gennes-Werthamer theory \cite{kane98}.
With increasing $x$, the $T_c(d_{\rm PdFe})$ curves are 
systematically lowered until for 
$x = 0.13$\, and 0.20 a nonmonotonic 
behavior with a shallow minimum between $d_{\rm PdFe}$ = 
10 - 20 \AA\, is seen. For these sample sets,
$T_c$ becomes independent of the interlayer 
thickness for large $d_{\rm PdFe}$. 
Note that for $x = 0.20$ all samples very likely have a ferromagnetic 
interlayer since already for $d_{\rm PdFe}$= 9 \AA\, 
a ferromagnetic hysteresis loop and a Curie 
temperature $T_{\rm C} \approx $\,70 K were measured (Table I).
Therefore, a nonmonotonic behavior caused by the establishment of 
long-range ferromagnetic order beyond
a certain interlayer thickness, as observed in Fe/Nb/Fe triple layers 
\cite{mueh96}, can be ruled out. 
 
The thickness where the minimum occurs for $x$ = 0.13 and 0.20 
is equal to the value 
of $d_c$ determined from the critical-field behavior. This indicates 
that once the films are decoupled the transition temperature 
becomes independent of the thickness of the ferromagnetic interlayer. 
This has also been found in V/Fe multilayers \cite{wong86,koor94}. 
For higher concentrations $x$ = 0.40 and 1 this behavior does not change 
qualitatively although the shallow minimum 
is no longer present. 
We note that $T_c$ still 
decreases for $d_{\rm PdFe} \geq d_c$\, although these
samples are already completely decoupled. 
Besides the dependence of $T_c(d_{\rm PdFe})$ a general decrease of 
$T_c(x)$ with increasing $x$ can be inferred when samples of 
nearly identical $d_{\rm PdFe}$ are compared. 
For $x$ = 0.13 and 0.20 a pronounced maximum of $T_c(d_{\rm PdFe})$ 
at larger $d_{\rm PdFe}$\, which would be attributed  
to a $\pi$-coupling mechanism as predicted by theory \cite{rado91}, 
is not observed. If such a mechanism does exist at all the 
absence of a $T_c$ enhancement can have several reasons. 
First, the transparancy for conduction electrons 
at the $S/F$ interface plays a crucial role 
\cite{rado91,khus97}. In the particular case of specular 
reflection at the interface 
the transparency is given by a parameter $\eta$ which 
is related to the ratio of the respective conductivities 
$\eta = \sigma_F/\sigma_S$ where any spin-dependent scattering 
at the $S/F$ interface is neglected. 
Applying the theory of Radovi\'{c} et al. \cite{rado91,buzd91} the overall 
$T_c(d_{\rm PdFe})$ dependence for $x = 0.20$ 
can be described with $\xi_F = d_c \approx $ 12 \AA\, and the parameter 
$\epsilon$ = 10.5 (Fig. \ref{fig8}, inset). 
(For comparison with $S/F$ multilayers, where the superconducting 
layer of thickness $d_s$ is in proximity 
with a ferromagnet at both boundaries, 
$d_S$ was taken $d_S$ = 2 $d_{Nb}$ 
= 400 \AA\, for the present case of $S/F/S$\, triple layers \cite{stru94}.)
Using $\xi_0 = 2 \xi_0/\pi$ = 57 \AA\, we obtain $\eta = \xi_F/\epsilon \xi_S 
\approx 0.02$. 
Similar low values have also been 
reported for Nb/Gd ($\eta$ = 0.047, 0.013) \cite{stru94,jian95} 
and Fe/Pb ($\eta$ = 0.04) 
samples \cite{gari98}, where the values were found to 
be much smaller than the ones obtained from the conductivity ratio.
This shows that $\eta$\, should be taken as a phenomenological parameter 
only. Besides the influence of 
electronic parameters on $\eta$\, 
the different structural interface quality of sputtered 
samples compared to samples prepared by evaporation is important, too. 
Second, the nonmonotonic behavior 
is completely suppressed if the spin-orbit scattering is strong compared 
to the ferromagnetic exchange interaction \cite{deml97,gari98}. 
We conclude that from $T_c$ measurements alone and 
the absence of a $T_c(d_F)$ oscillation it is not 
possible to prove or dismiss the existence of a $\pi$-coupling mechanism. 
Additional experiments on, for instance, $S/F/S$\,Josephson junctions 
have to be performed in the future to look for unequivocal evidence 
for such a phenomenon.

\section{Discussion}
Obviously, the concentration dependence of the critical-field 
behavior and of the transition temperature can be separated 
into two different scenarios. For low concentrations $x \leq 0.05$ 
the reduced crossover temperature $(1-T_0/T_c)$ as well as $T_c$ gradually 
decrease with increasing $d_{\rm PdFe}$. For these samples the critical 
thickness $d_c$ is large ($d_c \approx$\,100 \AA) and the individual 
Nb layers are coupled via the Pd$_{1-x}$Fe$_x$ interlayer. 
From the magnetic measurements it is 
not possible to decide unambiguously
whether these small-$d_{\rm PdFe}$\, samples exhibit long-range ferromagnetic order. 
In contrast, for $x \geq 0.13$ the magnetic measurements show definitely 
that the interlayer is ferromagnetically ordered. 
In this case a clear transition from coupled to decoupled behavior 
in $B_{c2 \parallel}(T)$, where $T_c$ becomes independent of $d_{\rm PdFe}$, 
is found. This transition appears at a thickness 
$d_c \approx$\, 12 \AA, much lower than for $x \leq 0.05$. 
This crossover allows an experimental determination of the 
characteristic penetration depth of Cooper pairs in the ferromagnetic layer, 
$\xi_F$, a quantity that 
is usually not accessible in experiment.
In the theory of Radovi\'{c} et al. \cite{rado91} 
the minimum of the $T_c(d_{\rm PdFe})$ curve 
appears around $d_F \approx \xi_F$. 
Therefore, the critical-field measurements may serve 
as an independent method for the determination 
of $\xi_F$.
First, we consider the two extreme cases of a pure Fe interlayer ($x$ = 1) 
and a pure Pd interlayer ($x$ = 0). For a pure Fe interlayer 
we can calculate the parameter $I = 4 \hbar D_F /\xi_F^2$ 
from the measured $\xi_F \approx$ 12 \AA. 
For the estimation of $D_F = v_{\rm F} l/3$ = 2.7 ${\rm cm}^2$/s 
we take $l \approx \xi_F$ as 
the electronic mean free path and a Fermi velocity $v_{\rm F}^{\rm Fe} 
= 6.9 \times 10^7$ cm/s for $sp$ electrons \cite{yous86}. 
This yields $I \approx$ 0.5 eV as a lower limit 
in fair agreement with $I \approx$ 1 eV 
for bulk iron. For a pure Pd interlayer, which is presumably not in 
a ferromagnetic state, at least not for $T >$ 1.5 K, the application 
of the theory is questionable. An upper limit of 
$I \approx$ 80 meV can be estimated as before, 
which is unrealistically low, 
even for the $4d-4d$ exchange interaction in pure Pd. 
However, spin fluctuations or "paramagnons" with a characteristic 
energy of $E$ = 21 meV
are possibly pairbreaking in Pd \cite{zara87}. Moreover,  
tunneling experiments on Pd-Pb sandwiches have shown that  
Pd is not gapless down 
to a thickness of 500 \AA\, which indicates that the 
pairbreaking is much weaker than by magnetic impurities and in magnetic 
fields \cite{dumo81} thus leading to a large $\xi_F$. Hence,
for the extreme cases of $x = 0$ and $x = 1$ 
the experimentally determined values of $\xi_F$ can be reasonably explained 
by pairbreaking due to the presence of spin fluctuations and 
by the exchange interaction, respectively.
Concerning the samples with an alloy interlayer 
the interesting point is that $d_c$, i.e. $\xi_F$,
is more or less independent of concentration for $0.13 \leq x \leq 1$. 
Band-structure calculations of ordered fcc Fe-Pd 
alloys show that the average 
exchange splitting decreases by no more than a factor of two when the Fe 
content is successively reduced from Fe$_3$Pd to FePd$_3$ \cite{mohn93}.
The decrease is basically due to the weaker $4d-4d$ exchange interaction 
compared to the $3d-3d$ interaction. Furthermore, the Fermi velocity 
of Pd is smaller than of Fe, 
$v_{\rm F}^{\rm Pd} = 2 \times 10^7$ cm/s 
\cite{dumo81}. This suggests that with decreasing $x$ the 
accompanying decrease 
of $I$ is more or less compensated by a decrease of $D_F$. Besides, 
$\xi_F$ depends only weakly on $D_F/I$, viz. $\propto \sqrt{D_F/I}$.
This explains why $d_c \approx$ constant for $0.13 \leq x \leq 1$.
For low concentrations $x \leq 0.05$ the critical thickness is much larger. 
Since in this regime the interlayer is likely to be in the paramagnetic state 
the reduction of $T_c$ is presumably due to the 
well-known pairbreaking caused by spin-flip scattering first investigated in the classical work 
by Hauser et al. \cite{haus66}. This is further corroborated 
by the gradual increase of $\xi_F$ with decreasing $x$ (Fig. \ref{fig7}).
Hence, the large change of $\xi_F$ around $x \approx 0.1$ marks the transition 
from a paramagnetic to a ferromagnetic interlayer, similar to the previous 
reports on Nb/Gd \cite{stru94} and Nb/Fe systems \cite{mueh96} where such a behavior 
was found in dependence of the thickness of the $F$ layer.
This large difference between $\xi_F$ in both regimes demonstrates that the 
pairbreaking by the exchange interaction is much stronger than by 
spin-flip scattering.
Hence, the concentration dependence of $d_c$ 
extracted from the superconducting measurements 
can be considered as a magnetic response, although the magnetic properties 
of very diluted samples could not be measured directly. 

\section{Conclusion}
The occurrence of a 2D-2D crossover in the parallel critical field of
Nb/Pd$_{1-x}$Fe$_x$/Nb triple layers allows a determination of the characteristic 
penetration depth of Cooper pairs $\xi _F$ in the $F$ layer. 
With increasing $x$, 
this length changes abruptly in a small 
concentration interval due to a transition from a paramagnetic to a ferromagnetic 
interlayer leading to a change in the underlying pairbreaking mechanism 
while it is found to be more or less independent 
of $x$ in the ferromagnetic regime.  
It is reassuring that the strong difference in $\xi_F$ between $x$ = 0.05 
and 0.13 is reflected in the different $T_c(d_{\rm PdFe})$ behavior.
Although the present work presents a step forward in identifying 
the requirements for possible $\pi$-junctions in $S/F$ layers 
or multilayers, a decisive experimental test for this possibility must 
await physe sensitive measurements such as the Josephson effect.

\acknowledgements
We thank M. Kelemen for performing the SQUID measurements.
This work was partly supported by the Deutsche Forschungsgemeinschaft 
through Graduiertenkolleg "Kollektive Ph\"anomene im Festk\"orper".

\end{document}